\documentclass[12pt]{iopart}

\usepackage{iopams}
\usepackage{graphicx}

\begin{document}

\title[CQED with charge-controlled quantum dots coupled to a fiber Fabry-Perot cavity]{Cavity quantum electrodynamics with charge-controlled quantum dots coupled to a fiber Fabry-Perot cavity}

\author{Javier Miguel-S\'anchez$^1$, Andreas Reinhard$^1$, Emre Togan$^1$, Thomas Volz$^1$ and Atac
Imamoglu$^1$\\ Benjamin Besga$^2$, Jakob Reichel$^2$ and J\'er\^ome Est\`eve$^2$}
\address{$^1$ETH Z\"{u}rich, Institute for Quantum Electronics, Wolfgang-Pauli-Strasse 16, HPT G5, 8093 Z\"{u}rich, Switzerland}
\address{$^2$ Laboratoire Kastler Brossel, ENS, UPMC-Paris 6, CNRS, 24 rue Lhomond, 75005 Paris, France}
\ead{sanchez@phys.ethz.ch}

\begin{abstract}
We demonstrate non-perturbative coupling between a single
self-assembled InGaAs quantum dot and an external fiber-mirror based
microcavity. Our results extend the previous realizations of tunable
microcavities while ensuring  spatial and spectral overlap between the
cavity-mode and the emitter by simultaneously allowing for
deterministic charge control of the quantum dots. Using resonant
spectroscopy, we show that the coupled quantum dot cavity system is at the onset of strong coupling, with a cooperativity parameter of 2.
Our results constitute a milestone towards the realization of a high
efficiency solid-state spin-photon interface.
\end{abstract}


\section{Introduction}
The interaction between a quantum emitter and a single optical
cavity mode, termed cavity quantum electrodynamics (QED), has
enabled a number of key experimental advances in quantum optics,
including the observation of enhancement of spontaneous emission
\cite{purcell1946spontaneous}, demonstration of the photon blockade
effect \cite{Birnbaum2005blockade} and vacuum induced transparency
\cite{tanji2011vacuum}. The key requirement for the observation of
the heretofore mentioned phenomena is a large cooperativity
parameter ($C=2g^2/\kappa \gamma$), \cite{Lugiato:1984ky} which is attained if the square of the coupling
strength ($g$) between the emitter and the mode exceeds the product
of the emitter ($\gamma$) and the mode ($\kappa$) energy decay rates.
Minimizing the ratio of the cavity mode volume ($V$) to its quality
factor ($Q$) allows for maximizing $C$, provided that the emitter is
located at a maximum of the cavity electric field (spatial overlap)
and the resonance frequency of the emitter and the mode are
identical (spectral overlap).

The condition $C \gg 1$ has been achieved for a number of different
emitters and cavity designs, ranging from single atoms coupled to
Fabry-Perot cavities \cite{Birnbaum2005blockade,Koch2012correlations} or on-chip micro-toroids
\cite{Dayan2008photonturnstile} to super-conducting qubits in
coplanar waveguide resonators working in the microwave domain
\cite{Wallraff2004strongcoupling}. A technologically very relevant
all-solid-state cavity QED platform in the optical domain consists
of quantum dots (QD) coupled to nano-fabricated cavities. However,
for these integrated devices achieving spectral and spatial overlap
has been a major challenge. Even though techniques that overcome
these limitations using state-of-the-art nanotechnology methods have
been demonstrated, a flexible cavity design where large $C$ can be
achieved for every QD would greatly improve the prospects for novel
solid-state cavity-QED experiments.

In this Article, we demonstrate $C = 2$ by coupling excitonic
transitions of single self-assembled QDs to a hybrid cavity
structure which consists of a GaAs/AlAs-based distributed Bragg
reflector (DBR) mirror below the QD layer, and a curved fiber-end
mirror approached from the top. Spectral and spatial overlap in this
structure is achieved by moving the sample with respect to the fixed
top mirror. Previous attempts with QDs in similar hybrid structures
realized $C$-values well below one
\cite{cui2006hemispherical,muller2009coupling,barbour2011tunable}.
Besides the significantly enhanced cooperativity value in the
present setup, the main novel feature is the fact that our QDs are
embedded in a $p-i-n$ structure: by separately contacting the $p$
and $n$ layers electrically and applying a gate voltage, we achieve
full charge control of the QDs. This in turn opens up the path to
perform cavity-QED experiments where optical transitions address
given QD spin states. We thus demonstrate a fully tunable
spin-cavity-QED system requiring a minimum of technological steps,
together with fiber-coupled optical output, that can in principle
satisfy the high collection efficiency requirement of quantum
information processing protocols.

\section{Cavity QD coupling}

The dynamics of a two-level emitter, e.g. an excitonic QD transition,
coupled to a single cavity mode is accurately described by the so-called
Jaynes-Cummings model~\cite{jaynes1963comparison}. The Hamiltonian includes
a coupling term between the cavity and the emitter, which is characterized by a pulsation $g$,
also called coupling strength. This coupling is proportional
to the scalar product of emitter dipole and the intra-cavity electric field
generated by a single photon \cite{andreani1999strong}
\begin{eqnarray}
g=\left( \frac{1}{4\pi\epsilon_0} \frac{\pi e^2 f}{m_0 V_m} \right )^{1/2} \label{gformula}
\end{eqnarray}

Here, $f$ is the oscillator strength of the emitter transition, $m_0$
the electron mass, and $V_m$ the effective mode volume which is
defined as
\begin{eqnarray}
V_m = \frac{\int \epsilon_r(\mathbf{r}) \, |\mathbf{E}(\mathbf{r})|^2 \, d^3r }{\max |\mathbf{E}(\mathbf{r})|^2} \label{veff}
\end{eqnarray}
$\epsilon_r(\mathbf{r})$ is the dielectric function,
whereas $\mathbf{E}(\mathbf{r})$ denotes the intracavity electric field.
Equation (\ref{gformula}) gives the maximal value of the coupling parameter $g$ assuming that the emitter is located at the maximum of the electric field. A small cavity volume enhances the coupling.
For strong enough coupling, the transmission and reflection of the coupled system are significantly modified compared to the bare cavity at the condition that the emitter and the cavity can be brought to resonance.\cite{englund2007controlling,srinivasan2007linear} The
effects of the cavity-emitter coupling can be quantified by the cooperativity $C$.
In the large cooperativity limit, a significant portion of the emitted light ends
in the cavity mode which can be efficiently extracted, making $C
\geq 1$ a desirable operating range for many applications including
quantum information processing. It is also well known that $C \geq 1$ is required for
obtaining strong photon-photon interactions. Last but not least, the fidelity of
cavity mediated qubit-qubit interactions typically scale with $C$. To achieve $C \geq 1$ is challenging
and generally requires careful engineering of the coupling. A common technique
to increase $g$ is by reducing the cavity volume and positioning the QD at
the cavity field maximum.

Usually, epitaxial QDs are randomly distributed on the wafer
surface. Experiments have tried to achieve spatial matching by
defining a nano-fabricated array of cavities on top of the randomly
distributed QDs. Even though strong coupling could be observed this
way the chance for close-to-optimal coupling is rather low. More
sophisticated methods for relative alignment of QD and cavity mode
have been reported in literature. The first approach controls the
position of the QDs on the wafer during growth by nucleation sites,
\cite{schneider2012ga,mohan2010polarization} but the QDs grown this
way still lack the close-to-transform-limited linewidths of their
randomly distributed counterparts by about one order of magnitude.
In the second approach the cavity is written around single
pre-selected (randomly distributed) QDs which are precisely located on the wafer by SEM
metrology \cite{hennessy2004positioning} or optical spectroscopy
\cite{dousse2008controlled,thon2009strong, dousse2009scalable}.

Even though these approaches enable excellent spatial overlap, the
resonance frequencies of cavity and QD are generally different which
implies the need of post-processing tuning mechanisms to bring the
system into resonance. Besides irreversible fine-tuning by digital
etching \cite{hennessy2005tuning} an in-situ reversible tuning
mechanism for photonic-crystal cavities by means of
adsorbing$/$desorbing gas molecules was demonstrated
\cite{mosor2005scanning} and is now widely used in many
laboratories.  The hybrid cavity-QED approach we detail here stands in stark contrast to  these
earlier approaches, since it allows for a much more
straightforward spatial and spectral alignment with any of the QDs
on a given device.

\begin{figure}[h!]
  \centering
          \includegraphics[height=4.0in]{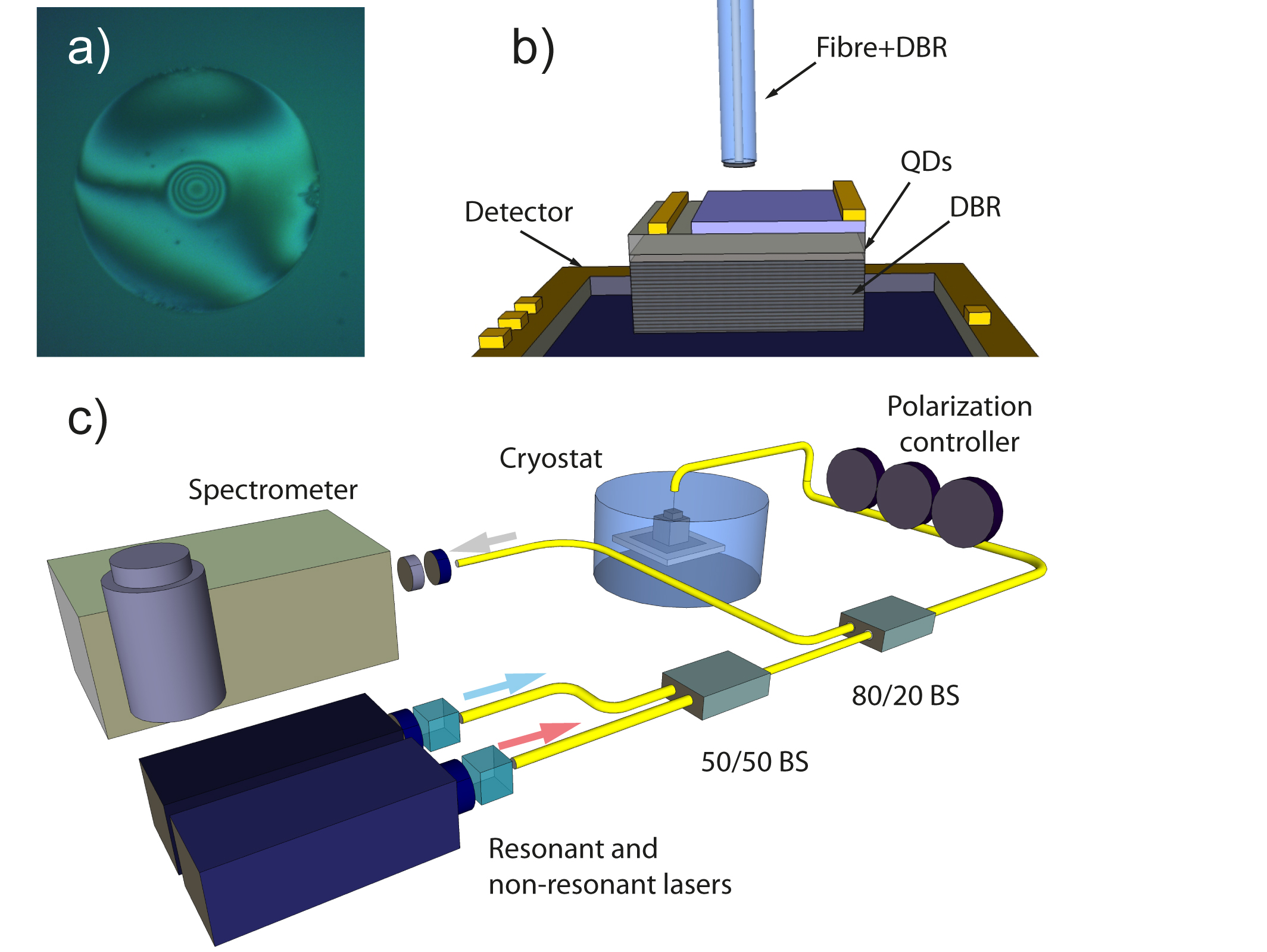}
                \caption{Setup of the semi-integrated QD-cavity system. a) The top mirror of
                the Fabry-Perot type cavity consists of a highly-reflective dielectric
                DBR mirror at the tip of a standard single-mode optical fiber. The central Gaussian recess
                leads to a denser set of interference rings in the profilometer picture shown.
                b) The planar bottom DBR mirror is made of 28 GaAs/AlAs layers with the active QD layer on top.
                The relative distance between fiber end and sample controls the cavity length and thereby
                cavity resonance frequency. c) Experimental setup. Off-resonant
                and near-resonant lasers are used to excite and probe the QD-cavity system by
                photoluminescence and transmission spectroscopy, respectively. Luminescence is
                collected through the same fiber that is used to excite the sample
                in the liquid-helium bath cryostat and analyzed on a grating spectrometer
                with integrated CCD chip. Fiber paddles control the polarization
                of the resonant laser light before it enters the cryostat.}
\end{figure}

\section{Experimental setup}

Our semi-integrated hybrid cavity system consists of a sample-based DBR
mirror below an active QD layer and a curved fiber-end mirror
(Figure 1a) which is approached from the top. The sample is mounted
on a stack of piezoelectric nanopositioners for precise
positioning in all three spatial dimensions. The fiber in turn is
fixed above the sample surface as illustrated in figure 1b). For
performing photoluminescence spectroscopy, an intensity-stabilised
pulsed Ti:Sapphire laser at 785nm is used. The system is excited
through the DBR-coated fiber mirror close to a reflectivity
minimum of the mirror. The photoluminescence is collected through
the same fiber and sent to a high-resolution grating spectrometer
with a nitrogen-cooled CCD camera for recording the spectrum. For
resonant spectroscopy, a mode-hop free intensity-stabilized diode
laser, tunable from 890 nm to 910 nm, is sent through the fiber, and
the transmitted light is collected on a silicon detector mounted at
the bottom of the sample (referred to as transmission detector
henceforth). The transmission signal is expected to be orders of
magnitude larger than the resonant reflection signal due to the
considerably higher reflectivity of the top fiber mirror (Transmission and losses for this mirror 
are 26 ppm and 13.5 ppm, respectively). The
transmission signal is directly amplified by a high-gain low-noise
amplifier. Sample and fiber are part of a home-built cage system
that is inserted into a buffer-gas filled dipstick which in turn
sits in a liquid He dewar and is kept cold at 4K.

\section{Fiber mirror cavity}
\subsection{Cavity modes}
To a good approximation, the cavity can be considered as a
planar-concave Fabry-Perot cavity whose properties are determined by
the radius of curvature $R$ of the concave mirror at the tip of the
fiber and by the length $L$ of the gap between the fiber and the
substrate. A given cavity mode is characterized by its polarization
and by three integers $q$ and $m,n$, which label the longitudinal
and the transverse mode structure, respectively. For a curved mirror
with rotational symmetry, one would expect the transverse modes with
equal values of $m+n$ to be degenerate. As figure 2a illustrates, in
our system modes with the identical values of $m+n$ are non
degenerate, because of the slight ellipticity of the fiber mirror.
In addition, all the modes exhibit a polarization splitting as
demonstrated for the TEM$_{00}$ mode in the inset of figure 2a. In
the remainder of this paper, we label the two orthogonally polarized
modes as TEM$_{00}^\mathrm{A}$ and TEM$_{00}^\mathrm{B}$.

\begin{figure}[h!]
  \centering
          \includegraphics[height=5.0in]{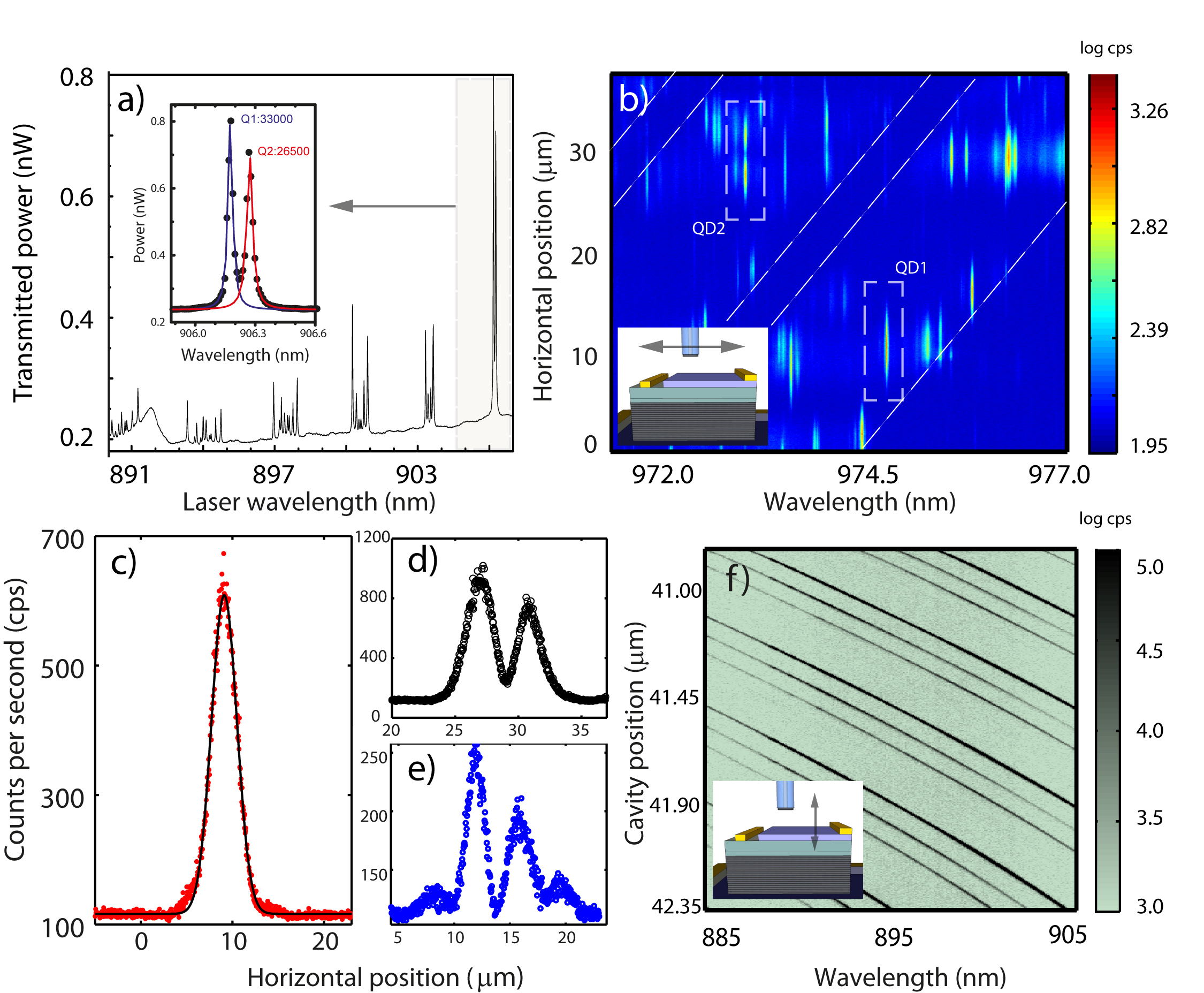}
                 \caption{a) Cavity transmission as a function of laser wavelength in the range between 890 and 907nm. The different groups of cavity resonances correspond to different transverse modes with the same
                 number of transverse excitations $n+m$. The resonance with the lowest energy is the fundamental TEM$_{00}$ mode which is split into two linearly polarized modes, TEM$_{00}^A$ and TEM$_{00}^B$. The splitting between these modes amounts to 144 $\mu$eV. b) Transverse mode profiles: By scanning the sample in the transverse direction, the mode profiles of the different transverse cavity modes are mapped out by recording the luminescence from the system as a function of position. Due to a slight wedge in the sample, the data were recorded using a slow modulation technique in the z-direction to always ensure coupling of the QDs to the cavity mode. This leads to "photoluminescence gaps" indicated by the tilted dashed white lines (see Appendix for details).  c) Gaussian fit (black) to the fundamental TEM$_{00}$ mode (red) measured in (b), with a $\omega_0$ of 2.7$\mu$m. d) and e) show higher order transverse modes profiles (QD3 not shown in scan (b)). f) Luminescence from the cavity modes as a function of cavity length demonstrating the tunability of the cavity resonance frequency. Due to off-resonant cavity feeding, the cavity modes are visible over a large range of wavelengths.}
\end{figure}

These effects are summarized by the following equation which gives
the resonance frequency of a mode

\begin{eqnarray}
\fl \nu_{qmn}^{A,B} = \frac{c}{4 \pi L} \left[ 2\pi q +
(2m+1) \arccos \sqrt{1-\frac{L}{R_x}} + (2n+1) \arccos
\sqrt{1-\frac{L}{R_y}} \right. \nonumber\\  + \left. \phi(\nu_{qmn}^{A,B}) \pm \delta \right]
\label{modefrequency}
\end{eqnarray}
The $A$ and $B$ symbols distinguishes the two eigenpolarizations which
correspond to two orthogonal linearly polarized modes. The
associated phase $-(+)\delta$ for the $A$ $(B)$ mode is a consequence of the small
birefringence present in both DBR mirrors and is on the order of
10~mrad. The two curvature radii $R_x$ and $R_y$ account for the
ellipticity of the fiber mirror (the two values typically differ by
a few percents), and the phase $\phi(\nu)$ is the sum of the phases
acquired through the reflection on each DBR. This phase varies
slowly around the common central wavelength of the DBRs and can be
considered as constant to first approximation ($\phi\approx\pi$ for
our sample), allowing a direct determination of the mode frequency.
In order to calculate resonance frequencies beyond this
approximation, equation \ref{modefrequency} must be solved with the frequency
dependence of the phase included explicitly which can be
determined using numerical methods, such as the transfer matrix method.

We expect the transverse distribution of the electric field in one
$m,n$ mode to be given by the corresponding Hermite-Gauss function.
This dependence can be observed by laterally scanning the fiber
above the sample and monitoring the emitted fluorescence collected
through the fiber on a spectrometer (see figure 2b). As the QD is
much smaller in size compared to the wavelength of light, and the
fiber only collects light in the cavity mode at the cavity
resonance, scanning QDs in space while exciting with a non-resonant laser
power above saturation
gives a very accurate measurement of the cavity mode profile.
Plotting the integrated fluorescence in a narrow frequency window
versus the lateral fiber position consequently gives a cut of the
intensity profile of this mode assuming only one QD to be present in
the narrow frequency window. Figure 2c-2e show three profiles
obtained from such measurements. As expected for the fundamental
mode, the profile fits quite accurately to a Gaussian (figure 2c).

The tunability of the cavity is assessed in figure 2f). By
collecting photoluminescence while slowly decreasing the cavity
length we observe how we can access smoothly many FSRs to couple the
QDs to any cavity energy for many different lengths. Emission from cavity modes are observed in a wide range due to the
fact that this particular scan was performed in a sample region with high QD density. The large variation of QD sizes is responsible for the broad band emission.
Being able to choose the cavity length while working with the same QD also opens up the possibility to build cavities with very long lifetimes. For example, the value of the quality factor $Q \approx 30000$ reported in the inset of figure 2a can be significantly increased (we have been able to measure $Q>60000$) if it is needed in a specific experiment.

\subsection{Cavity mode volume}
Considering a TEM$_{00}$ mode, the effective mode volume is given by
$\pi w_0^2 L_{\rm eff}\beta /4$ where $w_0$ is the mode waist on the
substrate, $L_{\rm eff}$ is the effective length of the mode
including the penetration depth of the cavity field into the DBRs,
and $\beta = |\mathbf{E}_{\rm vac}|^2 / \max |\mathbf{E}(\mathbf{r})|^2$ is the
ratio of the field maximum in the vacuum to the maximum in the whole cavity
(for our sample $\beta = 0.98$).
This length can be precisely estimated using a transfer matrix
calculation. For our sample, we obtain $L_{\rm eff} = L + 7.6$~$\mu$m.
The waist $w_0$ can be estimated from Gaussian optics to be
\begin{equation}
w_0=[\lambda^2 L(R-L)/\pi^2)]^{1/4}\label{spotsize}
\end{equation}
In order to minimize the mode volume and thus maximize the
cavity/emitter coupling, both $L$ and $R$ should be minimized. Using
the CO$_2$ laser ablation technique, radii down to 10~$\mu$m have been
reported. Here, we use a fiber mirror with a radius of curvature on the order of
75~$\mu$m \cite{hunger2010fiber}. The smallest $L$ that can be
achieved is often limited by geometrical aspects such as the depth
of the mirror structure at the fiber tip (on the order of a few
$\mu$m) and/or surface defects at the fiber tip. These are however
not fundamental reasons and cavities with lengths $L$ on the order
of a few wavelengths could in principle be fabricated. This sets the
minimal waist $\omega_0$ for this type of cavity to lie between 1 and
1.5~$\mu$m for a design wavelength close to 1~$\mu$m. In the
following, we minimize $L$ by moving the fiber down until it touches
the substrate and we step back by a few hundred nanometers. From the
coupling strength to the quantum dot that we measured (see section
8), we can estimate the effective mode volume to be smaller than
150~$\lambda^3$ for our current setup ($\lambda=$900~nm).

\section{The device: Charge controlled QDs}

A schematic of the sample is shown in figure 3a). It was grown by
molecular beam epitaxy (MBE) on a GaAs (100) substrate. The epitaxial structure growth starts with a 300 nm GaAs buffer, followed by 28
pairs of AlAs/GaAs that form the bottom mirror of the cavity, which sets the
timescale for the photon lifetime in the cavity since it has lower
reflectivity than the top fiber mirror. The reflectivity of the
bottom mirror was measured over the relevant wavelength range at
room temperature and is plotted in figure 2c). The interference
pattern below 860 nm is slightly smoothed out due to absorption by
the GaAs layers as was also confirmed by a transfer matrix
calculation.

\begin{figure}[h!]
  \centering
          \includegraphics[height=3.0in]{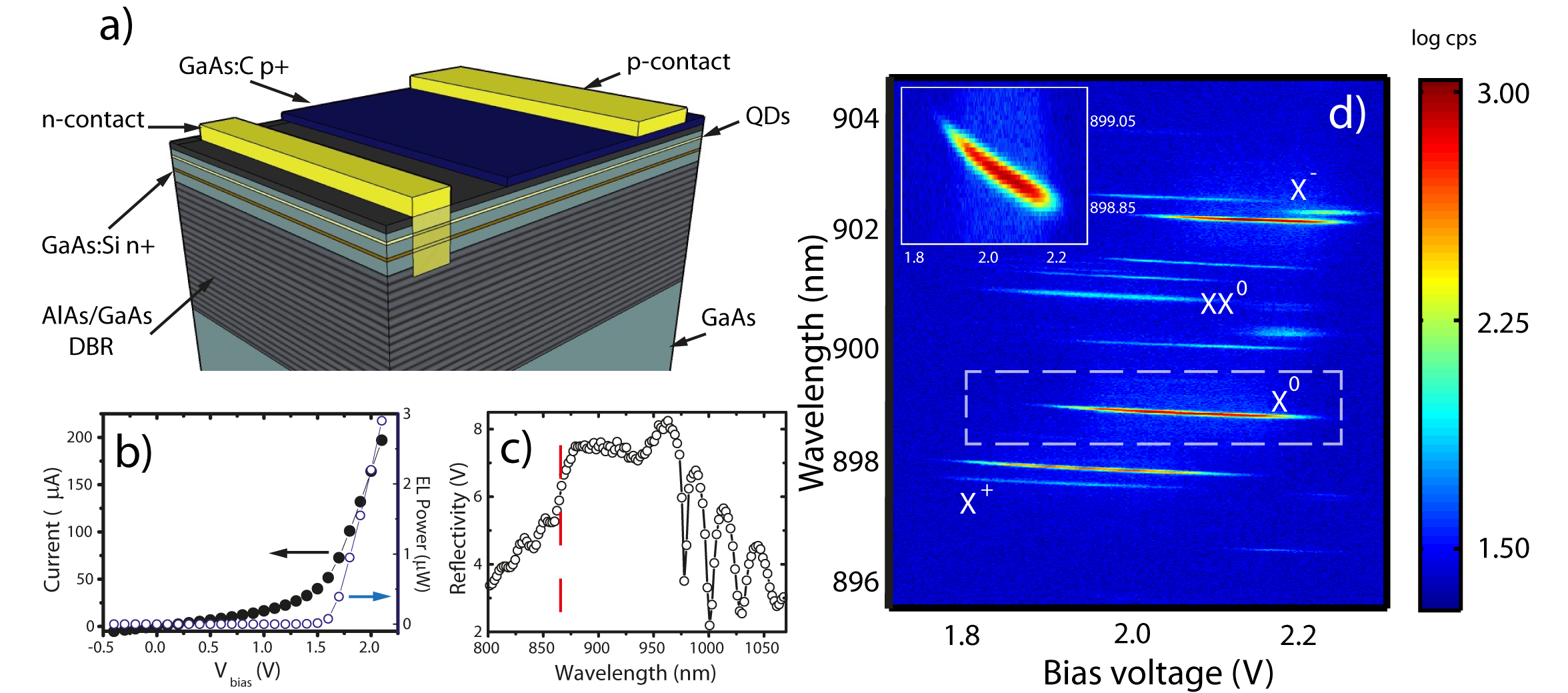}
          \caption{a) Device structure. In order to control the charge of the QDs, an n-doped GaAs layer below, together with a p-doped GaAs layer above the actual QD layer form a p-i-n diode structure. An additional AlGaAs blocking barrier between QD layer and p-doped region prevents excessive current flow. b) I-V characteristic (black) for the p-i-n diode and electroluminescence signal (blue) as a function of applied gate voltage at low temperature. c) Measured reflectivity of the semiconductor DBR mirror between 800 and 1050 nm. The vertical dashed red line indicates the GaAs bandgap energy. A minimum in reflectivity around 940 nm arises from a systematic measurement error. d) QD charging plateaus. Photoluminescence signal (logarithmic scale) as a function of gate voltage and wavelength. The different voltage regions corresponding to the different charging plateaus of the same QD are clearly visible. The inset shows a blow-up of the X$^{0}$plateau, with the quantum-confined Stark effect leading to a significant tilt in the plateau as a function of gate voltage. Note that for the measurement results shown here the sample was characterized in a flow cryostat leading to higher temperature, i.e. higher resistivity, and therefore to a higher voltage range than in part b)}

\end{figure}

A key feature of our system is the tunability of the exciton energy
and the control of the QD charging state. This is
achieved by a p-i-n structure on top of the AlAs/GaAs DBR. The
n-layer consists of a 40nm Si-doped GaAs layer with a carrier
concentration of $\approx$1$\times 10^{18}$ cm$^{-3}$, while the top
p-layer is a 35 nm wide GaAs layer doped with C atoms. The quantum
dots are sandwiched between the two conductive layers and were grown
without rotating the substrate to ensure a smooth gradient in QD
density across the wafer. The QD layer is separated from the
conductive n-layer through a 40 nm tunnel barrier of undoped GaAs. An additional AlGaAs blocking barrier
was introduced between the QD and the p-doped region. In the experiments
reported here, we used two different samples, with the QD emission
wavelengths centered at 900 nm and 970 nm, respectively. The emission
energy of the quantum dots was controlled by using the partially
covered island method \cite{garcia1997intermixing}. The overall
thickness of all the layers on top of the DBR mirror amounts to
$\lambda/n$, i.e. one optical wavelength.  This ensures that the QD
layer will be at an antinode of the cavity field along the growth
direction. The p- and the n-layers result in an intrinsic electrical
field that can be modulated by applying an additional external bias
voltage. For contacting the sample, the processing procedure was as
follows: In a small portion of the sample, the top $\approx$80 nm
were removed by wet etching with
H$_{2}$SO$_{4}$:H$_{2}$O$_{2}$:H$_{2}$O. In a second step, an Ohmic
contact to the n-doped layer was formed by annealing some Indium on
the sample surface for 360~s at an oven temperature of around
350$^\circ$C. The highly doped p-layer was contacted using silver
paint deposited on the sample surface. The I-V curve of the final
device is displayed in figure 3b). The deviations from the ideal
diode curve arise from the very simple processing protocol and unavoidable imperfections in the
sample due to microscopic structural defects. Due to the p-i-n
structure of the device, the application of a bias voltage induces
current flow of both electrons and holes which - through relaxation
into the QDs and the wetting layer - leads to spontaneous light
emission. Part of this electroluminescence signal was recorded on
our bottom detector when recording the I-V characteristics and is
plotted in the same figure 3b) (blue circles).

Recently, experiments based on photonic-crystal and
micropillar technology demonstrated charge control of QD excitons
\cite{laucht2009electrical,rakher2009externally,reitzenstein2011electrically,
pinotsi2011charge}, also in combination with resonant spectroscopy
\cite{pinotsi2011resonant}. While all of these approaches require
rather sophisticated fabrication procedures, it turns out that the
simple processing steps listed above are sufficient to obtain charge
control in our fiber-cavity setting. Charge control is demonstrated
in figure 3d: Here, photoluminescence emission was recorded as a
function of applied electric field. Different charging states can
clearly be identified in the spectrum, with the emission lines
originating from the neutral exciton (X$^{0}$), biexciton
(XX$^{0}$), and the positively (X$^{+}$) and negatively (X$^{-}$)
charged excitons (trions) indicated in the plot. In addition,
emission involving other multiply-charged states are visible. The
capability to deterministically charge the QD allows us to
selectively address trionic QD states and therefore make
use of the spin degrees of freedom. The spin
properties of the present sample were investigated (without fiber
mirror), and complete spin pumping in Faraday geometry
\cite{Atature2006spinpumping} was found. The inset of figure 3d)
displays the energy shift of the neutral exciton line X$^{0}$ due to
the quantum confined Stark effect as a function of applied electric
field. The significant Stark shift opens up the possibility of
electrically tuning excitonic states into resonance with the cavity
mode as will be demonstrated later in the article.

\section{Photoluminescence spectroscopy}

To demonstrate coupling of the cavity mode to single QD transitions,
we perform photoluminescence (PL) spectroscopy with an above-bandgap
pulsed laser at 785 nm. The emitted
photons from the coupled QD-cavity system are analyzed on a
grating spectrometer. In a first experiment, we fix the QD gate
voltage such that the X$^{0}$ exciton is stable. We continuously
collect PL spectra while tuning the cavity length and thereby
scanning the cavity resonance across the X$^{0}$ transition. The
result of this measurement is shown in Figure 4a. The cavity
resonance is detectable for all cavity lengths within the scanning
range even at very low powers due to off-resonant cavity feeding
\cite{winger2009explanation, yamaguchi2012third}. When the cavity
resonance wavelengths for the two non-degenerate orthogonally
linearly polarized modes TEM$_{00}^\mathrm{A}$ and
TEM$_{00}^\mathrm{B}$ match the X$^{0}$ emission wavelengths around
977.3 nm, there is a clear increase in the detected intensity. A
careful analysis, shown in Figure 4c, indicates that the detected
intensity follows a Lorentzian line shape as a function of cavity
length which is to be expected for an emitter weakly coupled to a
cavity.

\begin{figure}[h!]
  \centering
          \includegraphics[height=4.5in]{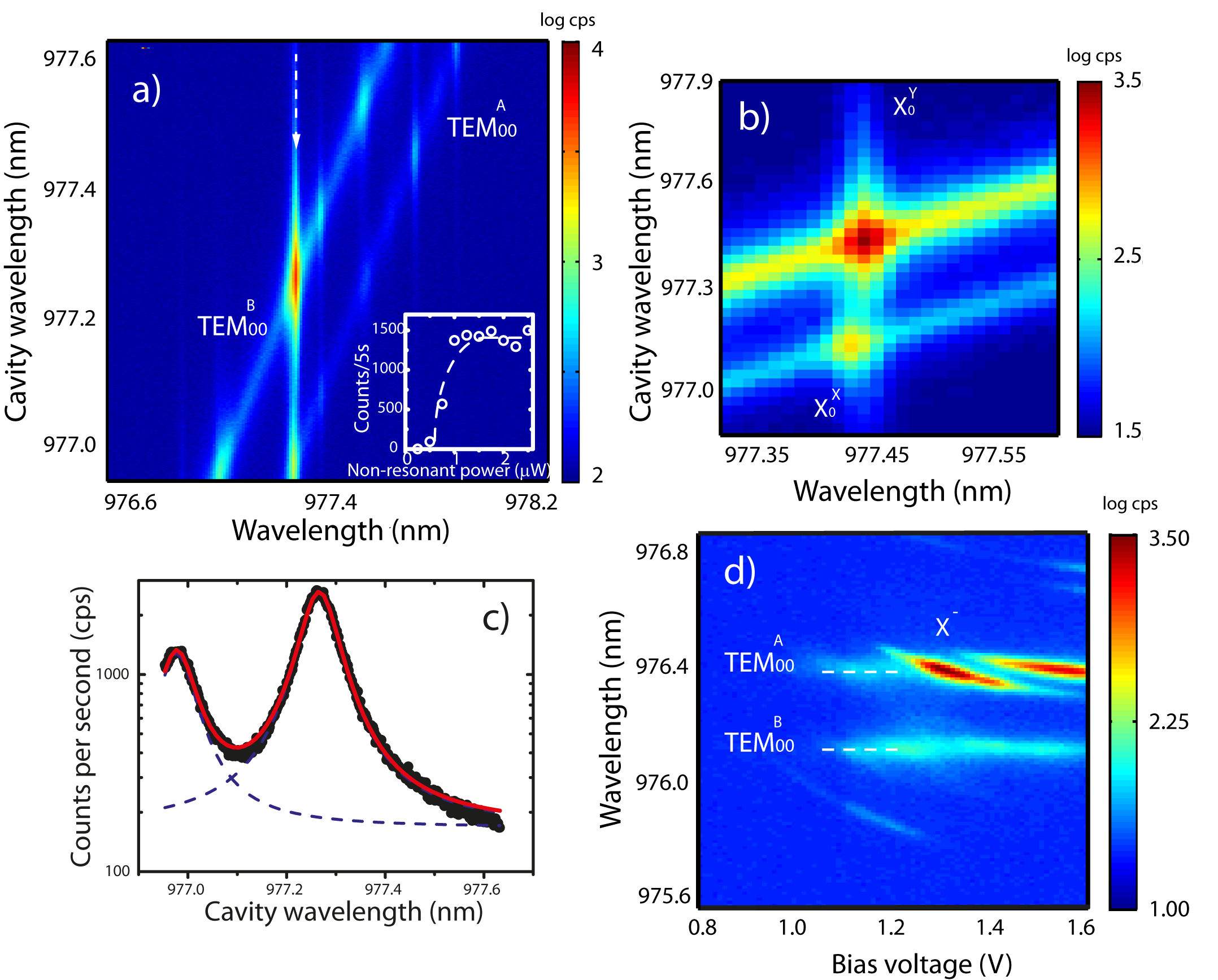}
                \caption{a) Tuning the cavity resonance wavelength through single quantum dot transitions. The data clearly show enhanced emission from the cavity-QD system when the cavity is resonant with single quantum dot transitions (vertical PL lines). The inset shows the saturation behaviour of the detected emission from the cavity-QD system as a function of off-resonant excitation power. b) XY-splitting of a neutral exciton. The two different polarization modes of the cavity (TEM$_{00}^\mathrm{A}$ and TEM$_{00}^\mathrm{B}$) couple preferentially to one of the neutral exciton transitions. The coupling strength depends on the relative angle between quantum dot axis and cavity polarization.c) Vertical cut of figure a) showing PL as a function of cavity
detuning (length). The data (black dots) is fitted by a Lorentzian
(red line) showing the characteristics of a weak coupling of the QD
to the cavity mode. d) Voltage tuning of single quantum dot lines to
cavity modes. Here, we change the bias voltage in order to tune
quantum dot transitions via the quantum-confined Stark effect into
resonance with the fundamental modes of the cavity. On resonance,
the QD emission is clearly enhanced.}
\end{figure}

As Figure 4b illustrates, for other QD exciton emission line, the wavelength at which the maximum
intensity is detected differs for the TEM$_{00}^\mathrm{A}$ and
TEM$_{00}^\mathrm{B}$ mode. We attribute this difference to the X-Y
splitting \cite{gammon1996fine} of the transitions of the $X^{0}$
charge state. For a single
QD, the coupling strength of a particular transition to a given
cavity mode depends on the relative angle between the QD axis and
the direction of linear polarization of the mode. Thus a rotator in
the system would allow for maximizing the QD-cavity coupling by
aligning the QD dipole emission along the cavity polarization axis.

A complementary PL spectrum can be obtained by fixing the cavity
length and tuning the QD bias voltage, which in turn tunes the QD
transition wavelengths via the quantum confined Stark effect. Such a
spectrum is shown in figure 4d). In this case, two spectrally close
excitonic emissions, which we tentatively attribute to $X^{-}$ and $X^{2-}$ are tuned in resonance
with one of the cavity modes. Again, the detected intensity
increases by an order of magnitude when the QD transitions are tuned
to the cavity resonances.

We also performed PL saturation measurements (pulsed excitation) with fixed cavity
length and bias voltage, with the cavity wavelength tuned to a
single QD excitonic transition. The intensity of the emitted light
from the QD-cavity system as a function of excitation power is displayed in the inset of
figure 4a). The clear saturation behaviour above a certain
excitation power provides evidence that one and only one QD is
coupled to the cavity mode for the particular wavelength detected
here.

A real door-opener for a multitude of future experiments with our
new system is its flexibility, reversibility and speed of tuning by
changing either the cavity length or controlling the gate voltage,
in particular when compared to more established techniques in other
systems such as gas deposition or temperature tuning
\cite{mosor2005scanning, winger2008quantum}. Another advantage of
the present system is the built-in fiber coupling which allows for
straightforward efficient probing using more sophisticated
techniques such as high-resolution resonant spectroscopy.

\section{Resonant spectroscopy}

We performed resonant laser spectroscopy by measuring the transmission
of a tunable diode laser through the coupled QD-cavity system using
the transmission detector. Since the top fiber mirror
has a higher reflectivity than the one grown on the sample, most of
the intra-cavity photons escape through that bottom mirror onto the
photodiode. The transmission signal can be directly detected with a
low-noise amplifier without the need for a lock-in technique.

\begin{figure}[h!]
  \centering
          \includegraphics[height=4.0in]{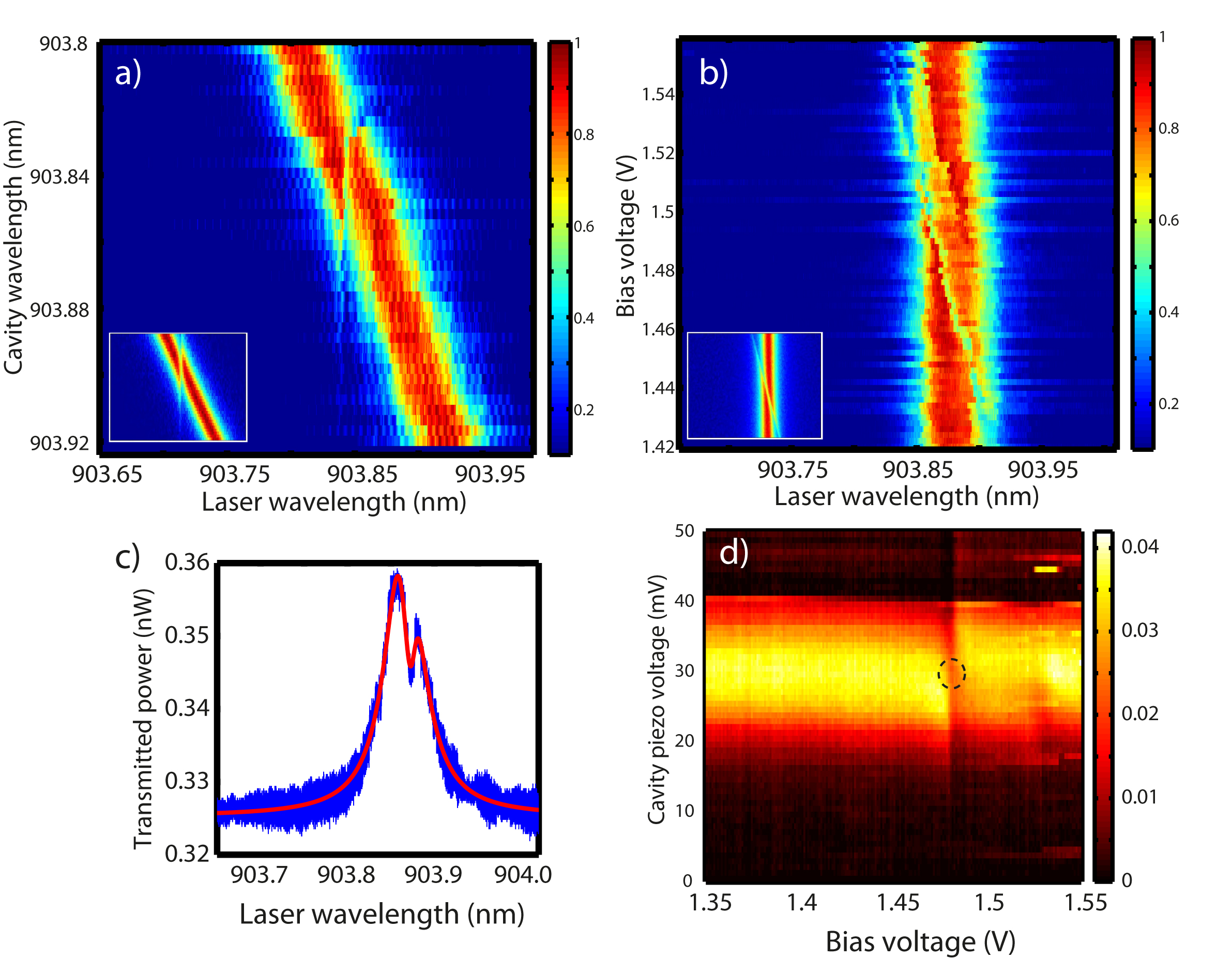}
                \caption{a) and b) 2-dimensional color plots of the system transmission normalized to maximal transmission while scanning the resonant laser (horizontal axis) as a function of a) cavity length and b) gate voltage. In a) the wavelength of the quantum dot transition stays constant whereas in b) the cavity resonance frequency is fixed. In both cases the dip due to the coupling of the QD to the cavity mode is clearly visible where one expects the QD resonance in the spectra. The two insets display corresponding calculated transmission spectra based on model, and parameters, used in text. c) Single transmission spectrum from b) at a bias voltage of 1.458 V. The contrast of the transmission dip is about 22$\%$. d) Keeping the laser wavelength fixed the system can be brought into resonance by a clever choice of both the cavity length and gate voltage. The sweet spot of maximum coupling is marked by the black circle.}
\end{figure}

Figure 5a) shows the transmission (normalized to the maximum transmission) spectra as a function of resonant
laser wavelength (horizontal axis) and cavity length (vertical
axis). The QD modifies the Lorentzian transmission of the
unperturbed cavity (top and bottom of the scan) by introducing a
dispersive response fixed at the QD resonance. As we performed all
the experiments well below saturation, we are able to extract the
coupling parameters directly from a fit to the data using \cite{AuffevesGarnier:2007el}
\begin{eqnarray}
T = T_0 \left |\frac{\kappa/2}{\omega-\omega_{\rm cav} + i \kappa/2 - \frac{g^2}{\omega -\omega_{\rm QD} +i \gamma/2}}  \right |^2 \label{transmission}
\end{eqnarray}

where $\omega_{cav}$ is the cavity angular frequency, $\omega_{QD}$
is the QD transition angular frequency and $T_0$ is the bare cavity transmission. From the experimental data we extract
$\hbar g$=11.7$\mu$eV, $\hbar\gamma$=11.4$\mu$eV, and $\hbar\kappa$=78$\mu$eV. The inset
in figure 5a \& 5b shows a calculated spectral map using these
parameters and the above expression \ref{transmission}.

In figure 5b, the cavity length is kept constant while the bias
voltage tunes the QD resonance through the cavity resonance. Note
that for each spectrum the background electroluminescence (fig 2b)
seen by the transmission detector was subtracted and finally
normalized to the peak transmission. Again as in figure 5a, the QD
significantly modifies the transmission spectrum through the cavity.

We plot the equation \ref{transmission} using the the previously extracted parameters in the inset of figure 5b). Figure 5c shows a horizontal line cut taken from data presented in Figure 5b at a bias voltage of 1.458 V (blue line), and its corresponding fit (red line). The dip appearing in transmission in Figures 5a-c when the laser is on resonance with the exciton is a spectacular signature of quantum
interference effect.~\cite{TanjiSuzuki:2011dk}

The great flexibility of our system is best illustrated by the fact
that one can demonstrate the QD-cavity coupling with an almost
randomly chosen laser wavelength by adjusting cavity length and QD
bias voltage simultaneously. We demonstrate this by parking the
laser at a wavelength of 903.865~nm and scanning both cavity length
and gate voltage. The resulting 2D map is displayed in figure 5d)
with the resonance condition indicated by the black circle.

\section{Onset of unity cooperativity and strong coupling}

A particulary interesting regime for experiments in cavity quantum
electrodynamics is the regime of strong coupling where the photon
exchange between emitter and cavity mode is as fast or faster than
the photon decay from the system. In this regime, new eigenstates,
so-called polaritons, form. In general, the eigenenergies of the
coupled QD-cavity system can be determined from \cite{reitzenstein2011semiconductor}
\begin{eqnarray}
\omega_\pm=\frac{\omega_{\rm cav}+\omega_{\rm QD}}{2}-i\frac{\kappa+\gamma}{4}\pm
\sqrt{g^2-\left( \frac{\kappa
-\gamma}{4} - i\frac{\omega_{\rm QD}-\omega_{\rm cav}}{2} \right)^2} \label{eigenenergies}
\end{eqnarray}

\begin{figure}[h!]
  \centering
          \includegraphics[height=2.5in]{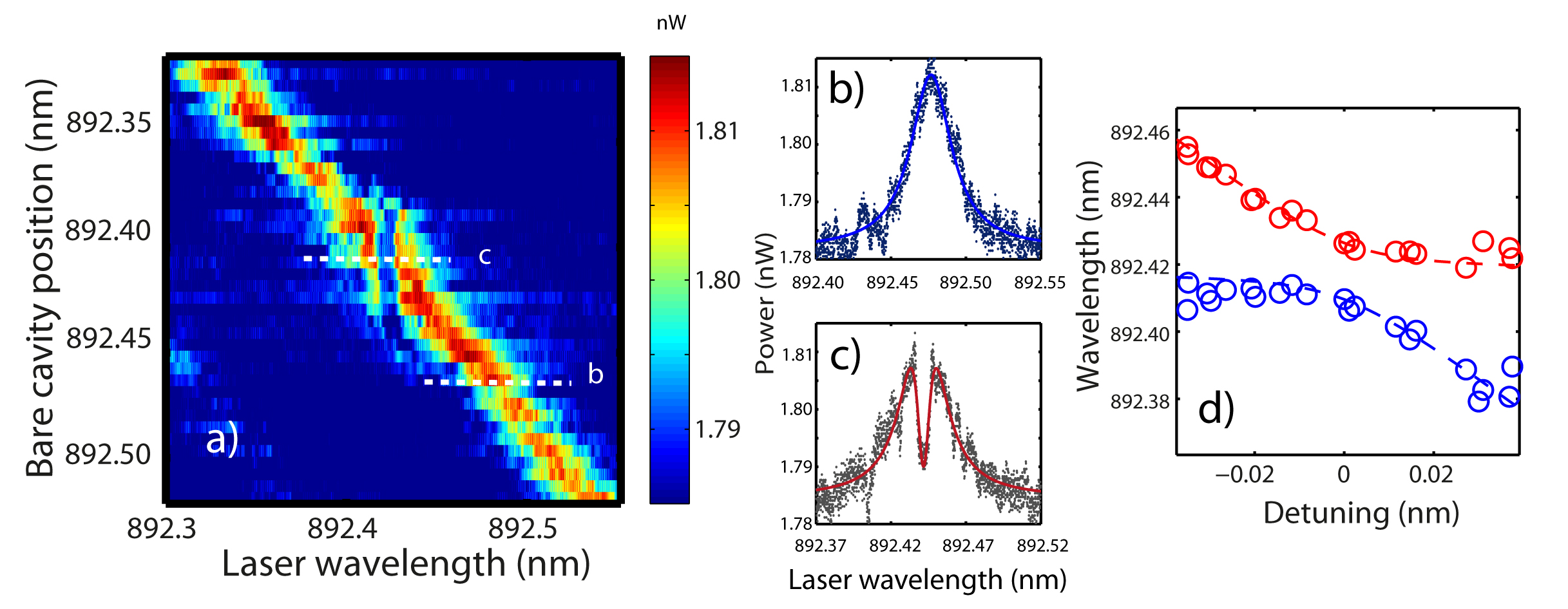}
                \caption{Onset of strong coupling a) Transmission as a function of cavity length and laser frequency.  Panels b) and c) display cuts off and on resonance. d) Using formula \ref{transmission}, the position of the two polariton modes was extracted as a function of cavity detuning. The dashed lines are the theoretical expectation for the g value extracted from the data in a).}
\end{figure}

By carefully positioning QD and cavity relative to each other, the
coupling strength can be optimized by reading out the PL counts.
Once an optimally coupled QD was
found, we recorded resonant transmission spectra as a function of
cavity length as displayed in figure 6a). Clearly, when crossing the
QD resonance, the cavity mode splits into two distinct peaks which
form an avoided level crossing. Fits to the spectra
(see e.g. figures 6 b) and c)) yield $\hbar g$=12.3$\pm$2.5$\mu$eV,
$\hbar\kappa$=50.0$\pm$2.7$\mu$eV, and $\hbar\gamma$=3.1$\pm$0.7$\mu$eV which
implies that for this particular QD with a narrow linewidth the system is at the onset of
strong coupling with $2g\approx (\kappa+\gamma)/2$.~\cite{Khitrova:2006ku} We fit the peak
positions for each spectrum in figure 6a) and plot the results in
figure 6d) which clearly yields an avoided level crossing that is
well described by equation \ref{eigenenergies} and the given value
for $g$.

The present QD-cavity system has a cooperativity of C$\approx$2.
In order to enlarge this number significantly, different strategies
can be adopted: increasing the number of layers of the DBR
semiconductor stack while simultaneously reducing the doping density
in the $p$ and $n$ layers should decrease $\kappa$ significantly. In
addition, a smaller radius of curvature of the fiber mirror would
reduce the effective mode volume of the cavity and hence increase
$g$. On the emitter side, other QD systems such as quantum-well
monolayer fluctuations with oscillator strengths up to one order of
magnitude larger \cite{peter2005exciton}, could push the system much deeper into the strong
coupling regime, thereby increasing the cooperativity up to an order
of magnitude.

\section{Conclusions and outlook}

In this article, we have presented a very versatile QD-microcavity
platform for performing state-of-the-art cavity QED experiments. The
system is fully tunable, i.e. both cavity length and QD energy can
be controlled at will. The high $Q$ of our system together with the
moderate mode volume brings us into the high-cooperativity regime
where the coherent interaction starts to dominate the system
dynamics. The onset of strong coupling was demonstrated through the
observation of an avoided level crossing in resonant transmission
spectroscopy. We anticipate that with some simple improvements on
both the cavity and emitter side, the system can enter deeply into
the strong coupling regime. The ability to control the charge state
of the QD by means of the $p-i-n$ structure will allow us to perform
experiments on quantum information processing with a first step
being the demonstration of an efficient fiber-coupled spin-photon
interface \cite{yilmaz2010quantum}. Our QD-fiber-cavity system might
then serve as a node in a future solid-state based quantum network
similar to what has been demonstrated with atom-cavity
interfaces \cite{Ritter2012quantumnetwork}.

\ack{The authors wish to thank F. Kaeding, S. Smolka and W. Wuester for
help in processing of the samples, M. Kroner for helpful discussions
and the Ohmic annealing oven, W. Gao and P. Fallahi for spin pumping
measurements and the ETHZ D-PHYS workshop for the fabrication of the
dip stick. We acknowledge support through NCCR quantum photonics, an
instrument of the Swiss National Science Foundation.}

\appendix

\section{Cavity modulation for QD spectroscopy}

\begin{figure}[h!]
  \centering
          \includegraphics[height=3.0in]{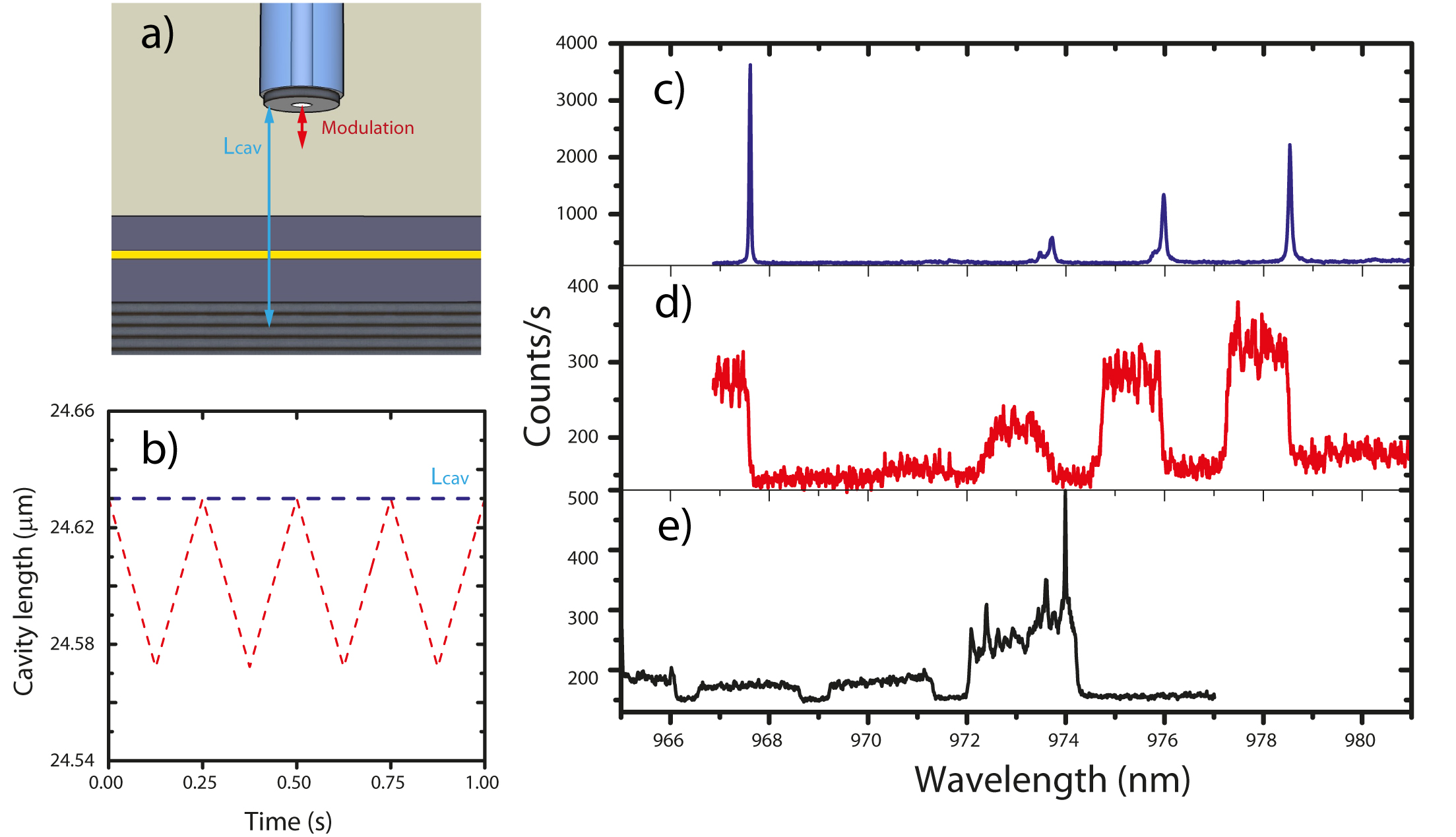}
                \caption{The principle of the modulated cavity is shown in (a): For a given cavity length we impose a triangle modulation (b) to create broader spectral windows through the cavity modes. Figure (c) shows the PL spectrum for a cavity length of $\approx$43$\mu$m integrated over 1s. The same spectrum broadened by the modulation (4 Hz, $\approx$58nm amplitude) is shown in (d). (e) shows the spectrum for a shorter cavity ($\approx$24.6$\mu$m) and the same modulation (5s integration time) when the cavity mode is spatially and spectrally aligned to a QD: The QD lines over the TEM$_{00}$ are clearly visible.}
\end{figure}

Sample is not always perfectly mounted flat due to fixing
imperfections, and it can have a slight angle with respect to the
fiber. Hence when performing horizontal scans with the fiber cavity,
the resonance of the cavity changes with horizontal position (for a
fixed vertical piezo position) due to the change in length of the
cavity. This makes some of the experiments requiring horizontal
scanning (determination of the mode profiles for example) more
involved, since the resonance of the cavity easily shifts away from
QD resonance. Here, we overcome this difficulty by modulating the
cavity length, $L_{eff}(t)=L_{eff}^0+f(t)$, with a triangular
waveform at a frequency $\omega_{osc}$. The amplitude f(t) thereby
varies between $f_{max}$ and $f_{min}$. This way, small
perturbations in cavity length do not matter anymore, provided the
the optical signal is aquired with a time constant $t \gg
2\pi/\omega_{osc}$. On the spectrometer, the narrow cavity modes are
transformed into broad rectangular resonances with a width given by
\begin{eqnarray}
\Delta \nu_{window}=(c/2n_{eff})(1/(L_{eff}^0+f_{min})-1/(L_{eff}^0+f_{max})) \label{window}
\end{eqnarray}
An example of such a spectrum is shown in figure A1 c). There we can
see the PL of a spectrum of the static cavity. As schematically
shown in figures A1 a) and A1 b), by applying a zigzag (triangular)
modulation of 4 Hz in frequency and amplitude of only $\approx$58nm,
we obtain the spectra seen in figure A1 d). There we can observe the
broad window through which we can detect the luminescence. As seen
in figure 3b) still by scanning horizontally the cavity the central
frequency of the cavity will shift with position (dashed lines in
figure 3b), but any QD measured within the modulated cavity window
can be analyzed (QD1 in figure 3b). An example spectrum of QDs
through the TEM$_{00}$ is shown in figure A e). Additionally, from
the slope of the dashed lines in figure 3b we could extract the
sample-fiber tip tilt of 0.15$^\circ$.

\section*{References}

  \bibliographystyle{unsrt}
      \bibliography{bibfile}

\end{document}